\documentclass[twocolumn]{emulateapj}

\newcommand{\apjvec}[1]{\mbox{\boldmath{$#1$}}}
\newcommand{\apjmat}[1]{{\mathbf{#1}}}

\newcommand{\vx}{\apjvec{x}}
\newcommand{\vy}{\apjvec{y}}
\newcommand{\vt}{\apjvec{\theta}}
\newcommand{\vk}{\apjvec{\xi}}
\newcommand{\mc}{\apjmat{C}}
\newcommand{\etal}{et~al.}

\begin{document}

\title{A Unified Framework for Photometric Redshifts}
\journalinfo{The Astrophysical Journal; received 2008 November 16, accepted 2009 January 14}

\author{Tam\'as Budav\'ari}
\affil{Dept.\ of Physics and Astronomy, The Johns Hopkins University, 3400 North Charles Street, Baltimore, MD 21218; budavari@jhu.edu}


\shortauthors{TAM\'AS BUDAV\'ARI}
\shorttitle{A UNIFIED FRAMEWORK FOR PHOTOMETRIC REDSHIFTS}

\begin{abstract}
We present a rigorous mathematical solution to photometric redshift estimation
and the more general inversion problem. The challenge we address is to meaningfully constrain unknown properties of
astronomical sources based on given observables, usually multicolor photometry, with
the help of a training set
that provides an empirical relation between the measurements and the desired quantities.
We establish a formalism that blurs the boundary between the traditional empirical and 
template-fitting algorithms, as both are just special cases that are discussed in detail to put them in context.
The new approach enables the development of more sophisticated methods that go beyond the classic
techniques to combine their advantages.
We look at the directions for further improvement in the methodology,
and examine the technical aspects of practical implementations.
We show how training sets are to be constructed and used consistently
for reliable estimation.
\end{abstract}



\section{Motivation} \label{sec:intro}

The concept of photometric redshift estimation is over four decades old. Since \citet{baum62}
the methodology has changed only incrementally but its role in astronomy has completely
spun around. The astronomy community originally received the idea with serious skepticism,
which, over time, thanks to a series of breakthroughs in the field \citep[e.g.,][]{koo85,connolly95a},
slowly faded. Today
the next generation telescopes plan to perform photometric observations only, and
completely rely on these kind of estimation techniques for most of their
key science projects including cosmology and large-scale structure.

While getting ready for extracting most of our new scientific knowledge
from photometric measurements,
we have to examine the current limitations of the various techniques and
understand the underlying assumptions. Essentially all currently existing
implementation can be
categorized into two classes of methods: empirical estimators
and template fitting.
Reviewing the history of the research area is outside the scope this study; see
\citet{weymann99} for a rich cross section of the field instead; now
we look at the basic concepts and the differences in the traditional methodologies.
Empirical methods map the relation of
the observed and desired properties using a training set; e.g.,
piecewise linear or polynomial fitting, or via other regression
methods like artificial neural nets, support vector machines, etc.
Template-fitting techniques rely on prior knowledge encoded in the model's
spectral energy distributions (SEDs) that can be matched to observations.
%
%
Why are the current implementations of these two so different?
There is no fundamental reason, e.g., one could generate training sets
from model templates.
Why do only template-fitting algorithms use photometric uncertainties
and not the empirical ones?
Why do people estimate the redshifts independently from other physical
properties, e.g., often use empirical redshift estimates and then
template spectra for type determination? We know these quantities are correlated and should be
dealt with in a consistent way.
The answers to these questions are usually direct consequences of
limitations in the models and the measurements.
If
the model SEDs matched all the observations, we would know everything
about all the objects in the Universe.
The uncertainties would be
used more often if they provided reliable extra information.

The ``Photo-Z'' label currently associated with the above methods,
should gain a new meaning.
We should expect more from the codes than a single estimate per object.
The implementations need to provide the full joint probability density
functions of all desired physical parameters, so we can develop 
new statistical tools that utilize all the information available.

In this paper, we are not concerned with what observables are the best to use or
which filter set is optimal for special cases of the generalized
photometric inversion problem, which depend on the specific science cases,
instead we derive a probabilistic formalism to address the common issues.
In Section~\ref{sec:method}, we introduce the methodology and derive the
formulas for determining the photometric constraints on physical
properties.
Section~\ref{sec:models}, describes the traditional empirical and
template-fitting algorithms as special cases of the proposed framework,
and the advanced techniques that go beyond their limits.
In Section~\ref{sec:disc}, we illustrate the concepts and detail the practical aspects.
Section~\ref{sec:sum} concludes our study.

Throughout the paper, we use the capital $P$ letter for probabilities and
the lower-case $p$ letter for probability density functions, or PDFs for short.

\section{Methodology} \label{sec:method}

We start by formulating the problem as general as possible.
The challenge is to constrain physical properties of sources
with some observables in a data set denoted by $Q$, hereafter the query set.
Since model spectra would never be perfectly suitable for all
desired parameters, one will need a training set, $T$.
In fact there is no reason to demand that these data sets have the
same observables. The mapping is provided by some model, $M$.
For example, magnitudes of different photometric systems
can be mapped on to one another, say, $U\!J\!F\!N$ observations to $ugriz$ \citep{fukugita95}
by empirical formulas.
In general, let $\vx$ be a set of observables in the training set $T$ that
also contains extra information about the physical properties $\vk$, and
let $\vy$ denote the observables of the query set $Q$. Our model is
parameterized by a vector $\vt$;
\begin{eqnarray}
T: \ & \ \left\{ \apjvec{x}_t, \apjvec{\xi}_t \right\}_{t \in T}  \nonumber \\
Q: \ & \ \left\{ \apjvec{y}_q \right\}_{q \in Q}  \nonumber \\
M: \ & \ \apjvec{\theta}  \nonumber
\end{eqnarray}
The model $M$ can predict the observables $\vx$ and $\vy$ for a given parameter via the
density $p(\vx,\vy|\vt,M)$ and has a prior on its parameters $p(\vt|M)$.
For example, one can build models based on the \citet{cww} or \citet{bc03} templates that
can be used to
calculate the colors of sources at a given redshift in any particular photometric system.
However, the modeling goes beyond just estimating the values for a given parameter, because
the observational uncertainties also enter the formula.
Later on, we will discuss in details how to establish various models; for now, the above
functions are assumed to be known.
Furthermore, let us assume that the training set samples the entire space of the
observables, and discuss the selection effects later.

Our goal is to derive the probability density function (PDF) of the physical
properties $\vk$ for a given query point $q$ with $\vy_q$ observations using our model $M$.
This function, $p(\vk|\vy_q,M)$, is the solution
of the generalized photometric inversion problem and the subject of this section.
The next two paragraphs discuss probabilistic concepts analogous to
elements of template fitting and empirical estimation,
respectively, in the context of our probabilistic formalism.
Next we address the burning issues of selection effects and feasibility.

\subsection{Mapping the Observables}

The first step is to make the connection between the observables.
It can be done
formally by calculating the probability density of $\vx$ for the query point $q$.
We do this via the equality of
\begin{equation} \label{eq:xy}
p(\vx|\vy_q,M) = \frac{p(\vx,\vy_q|M)}{p(\vy_q|M)}
\end{equation}
where the right-hand side contains integrals of known functions over the model's parameter domain
\begin{equation}
p(\vx,\vy_q|M) =  \int\!\!d\vt\ p(\vt|M)\,p(\vx,\vy_q|\vt,M)
\end{equation}
and over $\vx$ for the marginalization
\begin{equation}
p(\vy_q|M) =  \int\!\!d\vx\ p(\vx,\vy_q|M)
\end{equation}
We see how this is superior to the techniques analogous to the traditional way.
The usual solution involves fitting for the best-match model parameter using, for example,
maximum likelihood estimation (MLE), and accepting that parameter at face value to derive the estimates.
Here, we consider all possible model parameters and add up their contributions.

We note that the above general mapping formula is
valid in case of improper priors, too, in the sense that the posterior is always properly
normalized to unity.
If one has no prior knowledge about the model parameters,
and wishes to use a noninformative prior, e.g., flat \mbox{$p(\vt|M)\!=\!1$}, formally
he/she is allowed to do so; see more on the priors later on.

\subsection{Physical Properties}

Next we establish the relation between the observable and the desired physical
parameters. The traditional way is to assume the properties of interest
to be a function of the observables.
Some of the existing methods utilize explicit functions such as a polynomial or piecewise linear,
while others use more obscure mappings such as a decision tree or an artificial neural net.
Conceptually, they are just assuming a fitting function
\begin{equation} \label{eq:fit}
\vk = \hat{\vk}(\vx)
\end{equation}
which is tuned to reproduce the elements of the training set as best as possible.
The problem with this assumption is that there is no guarantee that the same $\vx$
observables always correspond to the same $\vk$ properties. In fact, we know that degeneracies
are present in most data sets.
Clearly, the above assumption is an unnecessary restriction over the general relation
of $\vx$ and $\vk$ denoted by $p(\vk|\vx)$. In other words, the traditional model is
\begin{equation} \label{eq:fitpdf}
p(\vk|\vx) = \delta(|\vk-\hat{\vk}(\vx)|)
\end{equation}
using Dirac's $\delta$ symbol.

A better way is not to restrict the distribution arbitrarily to an unknown
surface but to leave the formula general.
We can establish the proper relation by observing the fact that
\begin{equation} \label{eq:zx}
p(\vk|\vx) = \frac{p(\vk,\vx)}{p(\vx)}
\end{equation}
The right-hand side is a ratio of two densities that (both) can be
estimated from the training set, e.g., using Voronoi tessellation or
kernel density estimation (KDE).

Having derived the above relation, one can compute the final PDF of
interest as the integral over the possible observables in the training set
\begin{equation} \label{eq:pz}
p(\vk|\vy_q,M) = \int\!\!d\vx\ p(\vk|\vx)\,p(\vx|\vy_q,M)
\end{equation}
When it is possible to accurately characterize this distribution by a Gaussian function
or some mixture model, one can compress the numerical results
into a few parameters.
When the PDF is unimodal, which is often not the case,
the expectation value should suffice for an estimate
\begin{equation} \label{eq:res}
\bar{\vk}(\vy_q) = \int\!\!d\vk\,\vk\,p(\vk|\vy_q,M)
\end{equation}
The above equation is similar to kernel regression \citep{kernelregression}
in case of using KDE, except it is
a generalization to incorporate the uncertainties in the data sets.

Photometric redshifts and other such properties are often
used in statistical studies for their availability for a large
number of sources, even though they provide relatively
loose constraints on individual objects.
The full PDFs of the sources are best suited to derive the
ensemble properties
of entire catalogs or even specific subsamples.
The distribution of the properties over a set
of measurements $Q$ is given by the average
\begin{equation}
p(\vk|Q,M) = \left\langle p(\vk|\vy_q,M) \right\rangle_{q\in{}Q}
\end{equation}
Hence there is no need for an extra deconvolution step
to recover the underlying distribution of the objects
in a sample, because their average PDF is exactly that.
A common example is the estimation of the redshift distribution
$dN/dz$ for various subsamples, say, at different distances.
When selection bias is not an issue for the scientific analysis,
e.g., lensing studies, one can even choose the subsets to optimize
the contrast of the averaged PDFs.

\subsection{Selection Effects}

The inherent limitations of a finite training set pose a serious problem
for any estimator, which is often neglected. Our formalism introduced earlier
is no exception, hence we now turn to examine the effects.
The selection function is the probability of a source, with observables
$\vx$ making it into the training set, $P(T|\vx)$. The region that
the training set can sample is the window function $P(W|\vx)$, which
takes the value of 1 where the selection function is nonzero,
and 0 otherwise. For example,
\begin{equation}
P(W|\vx) =  \left\{\begin{array}{c l}
           1 & \quad \mbox{if\ $V\!(\vx)<22$}\\
           0 & \quad \mbox{otherwise}\\ \end{array} \right.
\end{equation}
for a survey that has a magnitude limit of 22 in a $V$ band.

The selection function is expected to enter our method at two separate places:
the marginalization over $\vx$ and via the density estimates used for the
relation $p(\vk|\vx)$.
The former appears to be inevitable but causes problems only at the boundaries
of the selection criteria. If the integrand $p(\vx|\vy_q,M)$ in equation~(\ref{eq:pz})
vanishes within the integration
domain of the window function $P(W|\vx)$, the results are valid.
Otherwise the estimated PDF is biased in an unknown way.
The probability of $q$ being inside the window function
is the right indicator of the problem occurring
\begin{equation}
P({W}|\vy_q,M) = \int\!\!d\vx\ P({W}|\vx)\,p(\vx|\vy_q,M)
\end{equation}
When this probability is close to 1, the training set provides
good support for the photometric inversion problem,
but when the value is low, the query point
is known to be outside the regime of the training set.

The relation between the desired properties and the observables
is the other issue as it is only probed on the training set.
The relation as seen on the training set depends on
the true relation and the selection function via the equation
\begin{equation}
p(\vk|\vx,T) =  \frac{p(\vk|\vx)\,P({T}|\vx,\vk)}{P({T}|\vx)}
\end{equation}
If the selection function strictly depends only on $\vx$, 
we have
$P(T|\vx,\vk) = P(T|\vx)$ and find that the empirical relation is
identical to the true one on the selection domain. If the sampling frequency
is low, the measured relation is noisier and less robust numerically.

This is a critical point, which is worth emphasizing once again: the $p(\vk|\vx,T) = p(\vk|\vx)$ equality holds
only if $\vk$ does not influence
the selection in any way, not even indirectly via some hidden parameter.
A counter example is the common case of cutting on morphological parameters
in the selection function, while only considering the fluxes for $\vx$.
Another interesting consequence is that one cannot use only the colors to estimate, say,
photometric redshifts, if a magnitude cut was involved in the selection of the training set.
Yet another issue is cosmic variance, which might cause the relation to depend
on the position in the sky.
The solution in all cases is to revise the selection of the training set, if possible,
or to add the hidden observables into $\vx$,
and extend the model to include them.

\section{Models in the Traditional Limits and Beyond} \label{sec:models}

Previously we have hinted at how models can be constructed
but, until now, they have just been assumed to be known.
A model is a combination of the limitations in our
observations, both in the training and query sets,
and the parameterization of the observables.
From discussing the topic in the most general way,
we now turn to the practicalities of real-life astronomical observations.

Today the errors of extracted fluxes of photometric measurements
are independent
estimates of the uncertainties in the separate passbands. Typically,
Gaussian errors are assumed, and the catalogs would quote 1$\sigma$ values
for every source.
Analyzing the repeated observations in the Sloan Digital Sky Survey \citep[SDSS;][]{york}, 
\citet{scranton_covar} have shown that this simple picture is wrong,
and the off-diagonal elements of the covariance matrix are significant.
This is not surprising. One of the major components in the photometric uncertainty is the error in
the determination of the aperture. If the multicolor measurements share a common
aperture, e.g., SDSS model magnitudes that are best suited for colors, the
flux measurements will be inevitably correlated.
Thus an improved error model of the photometric observations
is described by a multivariate normal distribution,
$N(\vx|\bar{\vx},\mc_x)$,
with a mean of $\bar{\vx}$ and covariance matrix $\mc_x$.
The next generation survey telescopes that plan to visit the sources on multiple occasions
will be able to better determine the full covariance matrices from actual observations
to improve our understanding of the errors. 
Hence, for now it is general enough to consider error estimates that are
fully described by the covariances.

In this reasonable approximation, the $p(\vx,\vy|\vt,M)$ mapping is also a normal distribution
with a full covariance matrix that includes cross-catalog terms, if necessary, that
go beyond the calibration work on the individual catalogs. If the apertures are
locked together for better color determination, one has to obtain the dependencies
via a data set that contains sources with all $\vx$ and $\vy$ measurements.
However, when the processing pipelines are independent, one can assume
that the uncertainties in
$\vx$ and $\vy$ are also independent, and write a realistic $M$ as
the product of the two Gaussians:
\begin{eqnarray}
p(\vx,\vy|\vt,M) & = & N_{x}\left(\vx|\bar{\vx}(\vt),\mc_{x}(\vt) \right) \nonumber \\
 & \times & N_{y}\left(\vy|\bar{\vy}(\vt),\mc_{y}(\vt)\right)
\end{eqnarray}
The dependences in the means $\bar{\vx}(\vt)$ and $\bar{\vy}(\vt)$ are straightforward to
model and, even in the most complicated case, are similar, in spirit, to the
traditional template-fitting procedures. For example, when considering a synthetic model of galaxies,
one has to vary the redshift, age, optical depth, and so on, to derive high-resolution
model spectra for different parameters, and then convolve them with the broadband filters
to get the fluxes.

Clearly modeling the covariance matrices is more complicated and would require
many more parameters to model accurately. If $\vt$ is a minimal set of parameters that
is enough to describe  $\bar{\vx}(\vt)$ and $\bar{\vy}(\vt)$, there are some other
hidden parameters or hyperparameters that are also needed for the covariances.
The fully Bayesian way is to establish
the relation of the covariance matrix and the hyperparameters
along with a hyperprior (the prior on the hyperparameters),
and to marginalize over the extra dependence.
Even though, this relation between the elements of the covariance matrix and the observables could,
in principle, be modeled based on the catalogs, it may prove impractical.
The empirical Bayes approach, admittedly more optimistic but easily quantifiable,
is to find the most likely hyperparameter and
substitute it into the dependence. In practice, for every parameter $\vt$,
one can find the values of $\bar{\vx}(\vt)$ and $\bar{\vy}(\vt)$ and
the closest measurement points, whose covariance matrices are good estimates.
If the covariance matrix changes slowly with $\vx$ compared to its
widths, one can safely calculate the values at the catalog points by using the
corresponding error matrices, 
\begin{eqnarray} \label{eq:normal}
p(\vx_t,\vy_q|\vt,M) & = & N_{x}\left(\vx_t|\bar{\vx}(\vt),\mc_{t} \right) \nonumber \\
 & \times & N_{y}\left(\vy_q|\bar{\vy}(\vt),\mc_{q}\right)
\end{eqnarray}
The only concern with this approximation is the noise on the elements of the
covariance. If needed, one could improve on the stability by smoothing or fitting
locally over the catalog entries.

The consequences of the model approximation in equation~(\ref{eq:normal}) are
most intriguing from the implementation
aspect of the methodology. As long as we only evaluate the PDFs at the observed
locations, the calculations are more straightforward and computationally less
expensive.

\subsection{Numerical Evaluation} \label{sec:num}

The field of numerical evaluation of complicated multidimensional integrals that
usually emerge in Bayesian analysis such as ours is well studied.
The solution
typically involves some randomized algorithms that range from simple
direct sampling from the prior to adaptive strategies often based on
Markov chain Monte Carlo (MCMC) methods, e.g., Gibbs sampling.
Although this topic is beyond the scope of the present discussion,
we briefly touch on the basic idea to illustrate the concepts
and provide some insight on how to derive
the final results numerically, namely
the value of $P(W|\vy_q,M)$ and the function $p(\vk|\vy_q,M)$.

The clever construction of the chain
in the MCMC algorithm yields model parameters $\{\vt_i\}$ that can be considered
independent random realization drawn from the posterior
distribution, $p(\vt|\vy_q,M)$ in our case.
With the chain in hand, one can readily approximate the integral
by the average over the MCMC samples. The mapping of the observables
then becomes
\begin{equation}
p(\vx|\vy_q,M) = \big\langle N_x\left(\vx|\bar{\vx}(\vt_i),\!\mc_t\right)\big\rangle
\end{equation}
where $t$ is the index of the training point $\vx_t$ closest to $\bar{\vx}(\vt_i)$.
When the query point is well within the regime of the training set, this
approximation is valid. What happens otherwise?
Often the uncertainties are larger outside the selection criteria, e.g.,
the photometric errors beyond the flux limit.
By using the covariance matrix of the closest training point, one
actually artificially decreases the contribution to the integral
making $p(\vx|\vy_q,M)$ tighter.
While the accuracy of the calculation is affected, the change is such that
it reduces the value of the integral in $P(W|\vy_q,M)$, which is the
measure of reliability. Hence, if we measure a large value, we can be confident
of the result.
Having said that we note that in practice the covariances probably do not change fast enough
to pose a significant problem in this calculation
for the objects along the edge of the selection function, and farther away
the probabilities are very small anyway.

Once we know that the estimation is in the safe regime, we can
compute the $p(\vk|\vy_q,M)$ integral ignoring the window function
completely by
summing up at preset $\vk_r$ points in our
region of interest, e.g., a fine redshift grid, as
\begin{equation} \label{eq:num_res}
p(\vk_r|\vy_q,T,M) \propto \sum_{t\in{}T} p(\vk_r|\vx_{t},T)\frac{p(\vx_{t}|\vy_q,M)}{p(\vx_t|T)}
\end{equation}
where the $p(\vx_t|T)$ densities and the matrix $p(\vk_r|\vx_{t},T)$ are
obtained from the numerical density estimates
once for the training set; see equation~(\ref{eq:zx}).
Here, we made use of the fact that the $\{\vx_t\}$ points are
(naturally) drawn from the distribution $p(\vx|T)$.

In order to perform these summations efficiently for many query points, one has to utilize
fast searching mechanisms in the space of the observables. The situation is complicated
by the strong correlation in the observables and the varying Mahalanobis metric, yet,
a significant speedup can be achieved by adequate multidimensional
indexing of the color--space as described in \citet{csabai07}.

\subsection{Template Fitting} \label{sec:template}

In classical SED-fitting approaches, one does not technically have a training set.
Although, formally it can be generated from a grid of model parameters $\{\vt_t\}$
as \mbox{$\{\vx_t,\vk_t\} = \{\bar{\vx}(\vt_t),\bar{\vk}(\vt_t)\}$}, where $\bar{\vk}(\vt)$ is
often simply a subset of $\vt$,
e.g., the redshift is just one of the parameters in the models of SEDs.
Traditionally, this artificial training set has no errors associated with the
reference points, hence we have
\begin{equation} \label{eq:diracx}
p(\vx|\vt,M) = \delta(|\vx-\bar{\vx}(\vt)|)
\end{equation}
and, assuming $\bar{\vx}(\vt)$ has an inverse,
\begin{equation} \label{eq:diracz}
p(\vk|\vx_t,M) = \delta(|\vk-\vk_t|)
\end{equation}
The analytical calculation yields an intuitive result,
where the grid points are weighted by their likelihood multiplied by the
corresponding prior
\begin{equation}
p(\vk|\vy_q,M) \propto \sum_{t\in{}T} \delta(|\vk\!-\!\vk_t|)\,p(\vt_t|M)\,N\!\!\left(\vy_q|\bar{\vy}(\vt_t),\!\mc_q\right)
\end{equation}
In the limit of a flat prior, this is
the classic MLE case, which is equivalent to the $\chi^2$ minimization
techniques used in most SED-fitting implementations today, where the measurements
are compared to the simulated observations at the grid points to select the optimum.
One obvious exception
is the algorithm of \citet{bpz}, which actually applies an explicit empirical
redshift prior in this equation, hence it is often referred to as ``Bayesian.''

The selection of a set of templates is another simple prior but on the spectral type,
even if well hidden, implicit, and not often admitted.
Researchers routinely seek for templates that provide the best redshift estimates.
Strictly speaking, this is cheating. The selection
should be based on how well the templates represent the data in the space
of the observables, and not based on their performance in the estimation.
Naturally, there is a connection, but not in that direction.
The templates that follow the data will likely provide better estimates;
however, templates that yield good estimates are not guaranteed to match
the data.
The development of a class of methods by \citet{budavari99,budavari00,budavari01}
and \citet{csabai00}
can be considered early attempts to achieve a better SED prior. Here, the templates are
statistically modified
to represent the observations more accurately, while not optimized for redshift estimation
whatsoever.
Clearly, these are just the first steps in this direction.
Instead of just assigning 1 and 0
weights to the templates by either including them or not (respectively) as
typically done today,
one can explicitly formalize more realistic priors over a broader range of
SEDs that are driven by scientific knowledge and/or ensemble statistics.

An obvious but rather important improvement in the new framework
is the ability to naturally introduce and utilize the
uncertainties of the template spectra. We know that the models are not perfect, and
this can be easily characterized. As an example, one can use the same prescription for
the spectral synthesis, but build on a various stellar libraries to
analyze the differences. When using empirical templates, the implementation is even more evident.
We fold in the uncertainties by abandoning the
simplified relation in equation~(\ref{eq:diracx}) and creating a more realistic model
with the estimated finite errors.

\subsection{Empirical Method} \label{sec:emp}

The new methodology in the limit of the classic empirical algorithms goes well
beyond the usual techniques, which consist of simply establishing the fitting
function in equation~(\ref{eq:fit}). We can utilize those fits (or preferably
estimate the densities numerically to map the full relation),
but we can also properly consider the uncertainties.

The parameterization of a minimalist model is done by a position in the space of the observables,
i.e., $\vt$ is the same type of quantity as $\vx$ and $\vy$, e.g., $U\!B\!V\!I$ fluxes.
Namely, we choose \mbox{$\bar{\vx}(\vt) = \vt$} and \mbox{$\bar{\vy}(\vt) = \vt$}.
Even though the observables in $\vx$ and $\vy$ are the same quantities,
the mapping is still required to fold in the photometric errors.
With an improper flat prior $p(\vt|M)=1$, the mapping of the observables is
integrated analytically
\begin{equation} \label{eq:emp_map}
p(\vx_t|\vy_q,M) = \int\!\!d\vt\ N(\vx_t|\vt,\mc_t)\,N(\vt|\vy_q,\mc_q)
\end{equation}
While this model is clearly very simple,
it is quite powerful and conceptually more sound than a number of traditional methods.
We will use it for illustrations in the upcoming discussions.

Other simple forms of priors can also be handled analytically, 
e.g., linear and Gaussian, that
may be reasonable approximations at least locally.
Otherwise we resort to the numerical evaluation.

\subsection{Advanced Methods of the Future}

The problem with the classic empirical methods is the requirement of having
the same set of observables for both the training and the query sets.
The limitations of the
SED-fitting techniques come from the fact that the models cannot perfectly
describe the relation of observables and the physical properties.

In the realm of our unified framework, we can have more advanced methods that
combine these two previously separate classes of techniques. We can introduce
new algorithms to
take advantage of the training points even if their photometric observables
differ from those in the query set.
The idea is the following: True to the spirit of empirical methods,
we utilize the training set to provide the
relation between the physical properties we wish to constrain and some
observables; see equation~(\ref{eq:zx}).
In addition to this empirical relation, we apply a mapping
from the observables of the query set to that of the training set
based on SED modeling, like in the template fitting procedures.
For example, if the training set contains
$U\!J\!F\!N$ magnitudes, one can map them to $ugriz$ using equation~(\ref{eq:xy}).

The intriguing observation to make here
is that one does not even need realistic physical models to start with,
because the physics is in the training set and not the model.
Let us consider a model $M$, which is a complete
basis on the observed wavelength range, e.g.,
Legendre polynomials or Fourier series with the parameterization
by their coefficients.
The manifold of the physical spectra is naturally contained within.
In practice, this model needs to be only sufficiently complete and band-limited
so that real SEDs can be well described; this is a weak prior that
we can set up based on all the spectra we observed and simulated before.
A model spectrum corresponding to a certain parameter value in $M$ can be convolved with the
appropriate transmission curves to yield the observables $\bar{\vx}(\vt)$ and $\bar{\vy}(\vt)$,
even if they are unphysical. Hence, formally we have the basis of our mapping, $p(\vx,\vy|\vt,M)$.
As long as the data provide
good enough constraints on the model parameters, the mapping is valid and
the algorithm follows the routine.
When the observations barely constrain the model parameters and large volumes
of unphysical SEDs have significant likelihood, the mapping will be wrong.
The solution is to apply a prior to consider only the physical SEDs.
Using the entire catalog, one
can derive an empirical physical prior statistically, which
we will discuss in the next section.

These new advanced methods overcome the usual difficulties in photometric redshift
estimation, and offer a way out of the half-century-old dilemma.
They are a natural extension of everything that has worked before in the field:
a straightforward combination of the two previously separate methodologies.

\begin{figure*}
\plotone{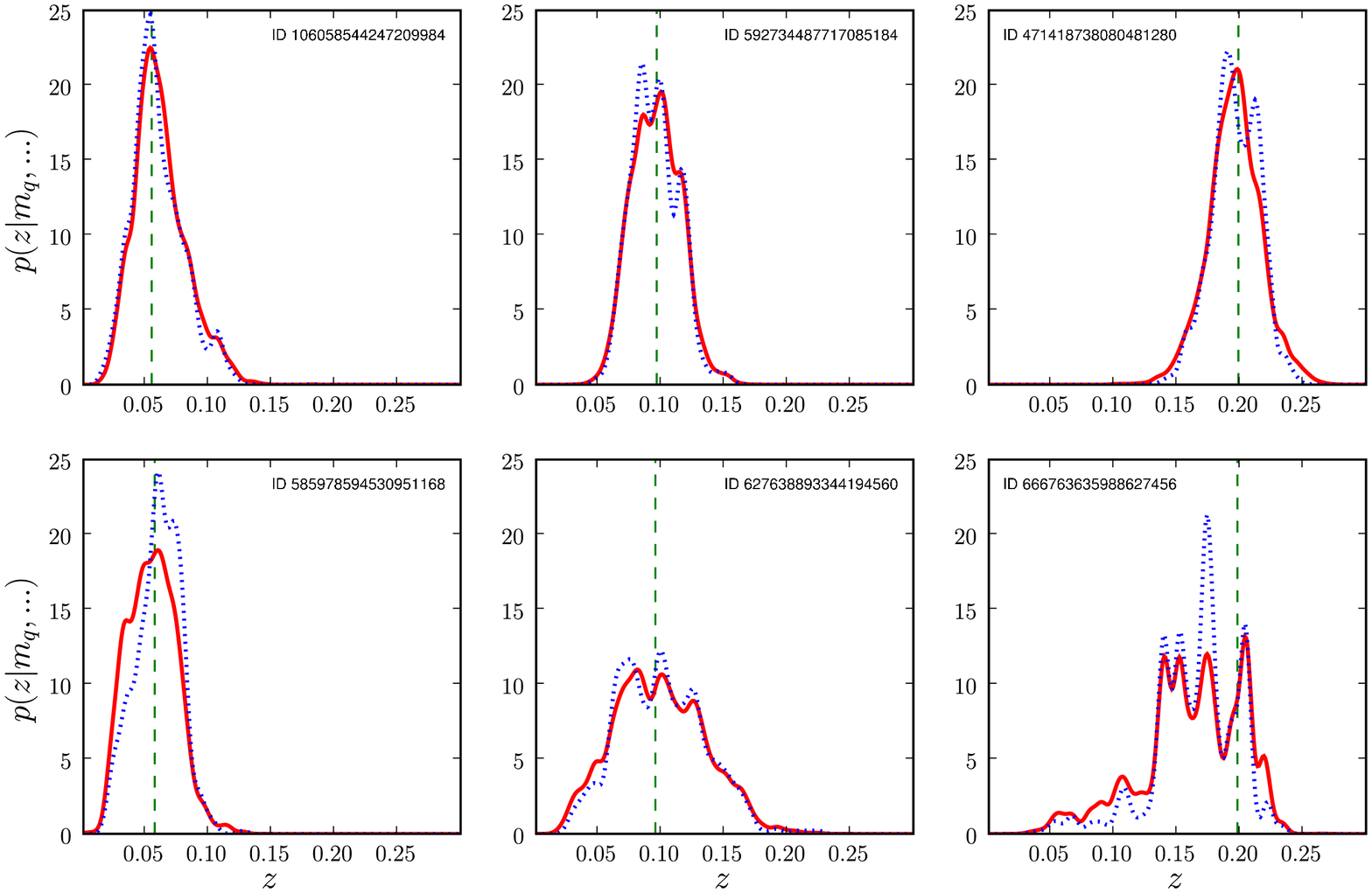}
\caption{The probability density as a function of the redshift for
early- and late-type galaxies ({\em{}upper} and {\em{}lower} panels, respectively)
at different distances marked by the vertical {\em{}dashed} lines. For
every object, the
{\em{}dotted} line shows the empirical relation of $p(z|\vx=\apjvec{m}_q)$,
and the
{\em{}solid} line illustrates the final result of $p(z|\vy=\apjvec{m}_q,M)$ after
properly folding in the photometric uncertainties via the mapping in the model.}
\label{fig:pdf}
\end{figure*}

\section{Discussion} \label{sec:disc}

Next we demonstrate the new framework in action by applying a simple
model to real-life data,
which is followed by discussions of the qualities of training sets and the prior.

\subsection{A Case Study}

To illustrate the concepts introduced earlier, we apply the aforementioned minimalist
empirical model to a sample of galaxies. We choose SDSS sources
for their well-studied photometric uncertainties. Following \citet{scranton_covar},
we estimate the full covariance matrix for all objects, and utilize them
in the subsequent analysis.
We randomly select a quarter of the entire Main Galaxy Sample \citep[MGS;][]{sdss_mgs}
of DR6 to be the
training set, roughly 100 thousand objects. Our query set is a smaller disjoint random
subset for illustration purposes.
First, we map the observables (magnitudes to magnitudes)
analytically using our simple model in equation~(\ref{eq:emp_map}). The calculation
is done inside the DR6 database by SQL User-Defined Functions.

Next, we compute the conditional PDFs by a dual-tree KDE implementation
\citep{gray03nde,lee06fgs}
at preset locations defined by the $T$ training and $Q$ query sets
in magnitude space and a uniform high-resolution redshift grid.
The practical complication with any density estimation is the fact that it is
scale-dependent and changes with the metric.
We are further limited in our applications to fix bandwidths for the conditional
density estimation in the current implementation of the estimator.
We adopt a bandwidth of $h=0.004$ in a metric that scales the magnitudes to the redshift.
In other words, the resolution in redshift space is set by $h$,
the full width half maximum of the normal distributions, and we re-scale the magnitudes
by a factor of $f=0.08$ to reasonably match the density of the sources in the
separate subspaces. This simple technique is expected perform reasonably well
within the regime where
the sources are suitably dense but not in the outskirts where a larger variable bandwidth is
needed in magnitude space. The theory of more sophisticated conditional
density estimation is well-studied \citep[e.g.,][]{fan96cde}, and advanced adaptive
implementations are in the works to help out (Lee \& Gray, 2008; private communication).

Figure~\ref{fig:pdf} illustrates the nature of the $\vx\!\!-\!\!\vk$ relation,
in this case the multicolor measurements and redshift $p(z|\apjvec{m}_q)$,
as well as the final redshift distribution, $p(z|\apjvec{m}_q,M)$, incorporating the photometric uncertainties
in our model.
We see that the redshift is really not a simple function of the magnitudes but
rather a more general relation. This is even more so for observables that constrain
the physical properties less than the $ugriz$ measurements.
The relation itself (shown as a dotted line)
might provide an overly optimistic view of the uncertainties at times
and usually much noisier than the final PDF (shown in solid)
that sums up these relations with appropriate weights.
The top panels show intrinsically red galaxies at three different redshifts,
which were selected based on the mixing angle
of the first two principal components, also known as the \texttt{eClass} in the SDSS
terminology. The bottom panels show the more problematic blue galaxies at similar redshifts.
Note the consistent performance of the estimator on the red sources
as a function redshift in comparison to the blue galaxies that have
broader PDFs at higher redshifts and are noisier, especially at the largest distances.

In the bottom rightmost panel, the distribution is not even centered around the spectroscopic redshift,
but skewed toward lower values. This object is very close to the edge of the training set,
and the result would be considered unreliable due to the
lack of calibrators at higher redshift that would still be within the sources
photometric uncertainties.

\begin{figure*}
\plotone{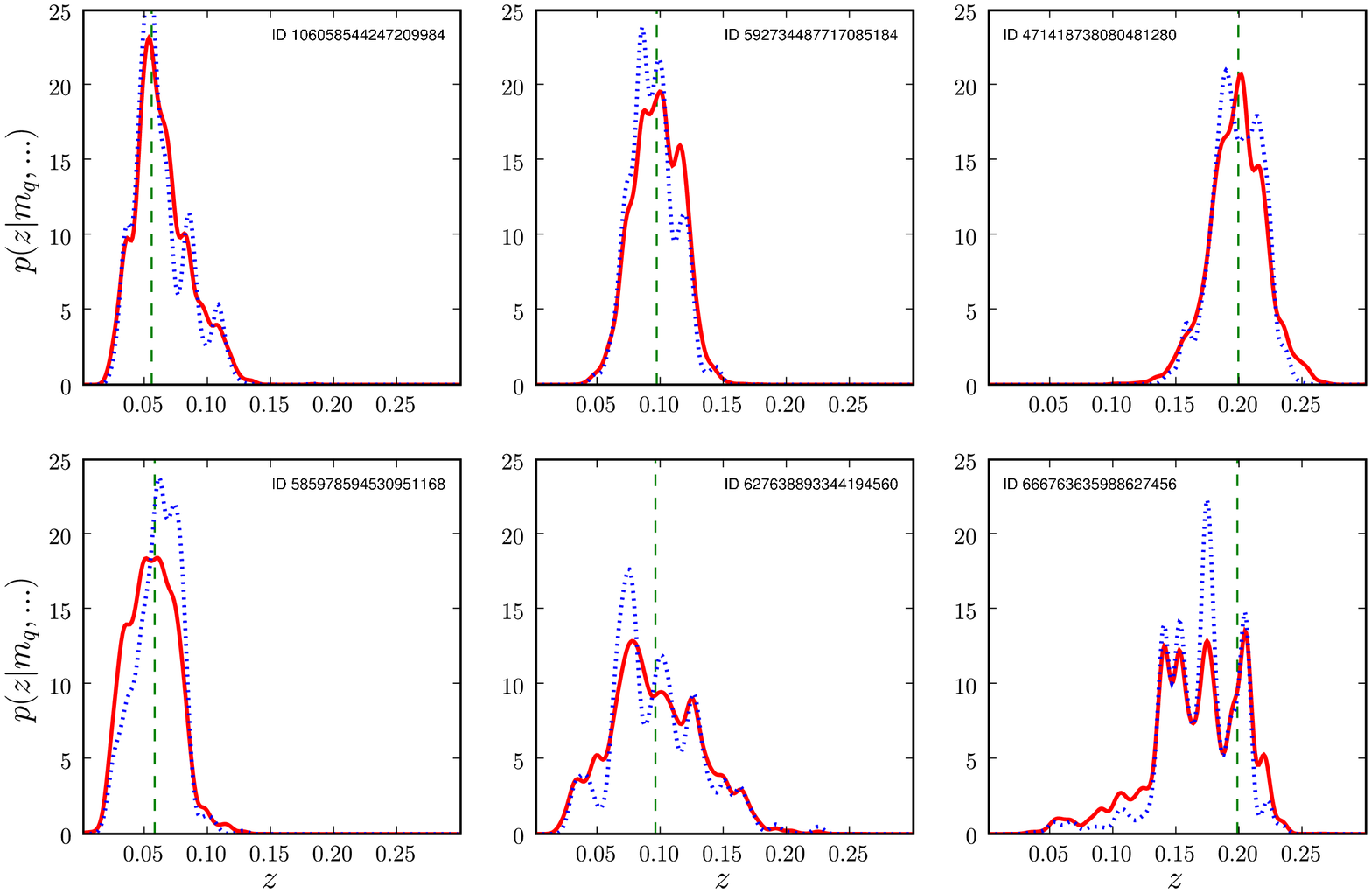}
\caption{Results obtained from the stratified training set look much like the
those from the full sample, which shows that the representativeness does not matter;
instead the training sets should be optimized for the broad coverage of the observable volume
with highest sampling rates in the outskirts.}
\label{fig:strat}
\end{figure*}

\subsection{Sampling Frequency}

A very attractive feature of the new PDF estimator derived earlier
in equation~(\ref{eq:res}) is its conceptual independence from the sampling
of the calibrators.
Many statistical tools rely heavily on having a representative training set and
only provide unbiased results in that limit. In our case,
the training points simply provide locations where the evaluation is feasible
and their density is essentially just a resolution factor.
The sampling frequency of the training set only affects the accuracy
of the numerical integral in equation~(\ref{eq:num_res}) but not in a systematic way
as long as the query point is well within the boundaries of the window function.
A denser training set will provide higher resolution in the summation,
but there is a practical limit beyond which one expects no improvement.
The reason is that the new calibrator sources are essentially identical to
the ones already in the training set.

The number of spectroscopic measurements to be carried out for calibration purposes
is limited by finite resources.
It is vital to acquire reliable training sets for the new generation
photometric studies.
A good training set has a well-defined selection function, using criteria
based on only the observables one plans to model for the estimation, and,
within that, a smart adaptive sampling strategy to optimize the coverage
in observable space. Clearly, the densest regions can be subsampled, but
one needs all training points in the outskirts of the manifold for broad
support.
For this reason, the simplest random subsampling of the underlying population
will not suffice. Instead, a stratified sampling strategy is to be pursued.

To demonstrate that the methodology is robust to this kind of systematic changes in the training set,
we create a stratified subset and perform
the previous analysis the same way. The sampling is done by including sources randomly
based on their local density $p(\vx|T)$ in magnitude space. A galaxy is included
in the training set only if the ratio of some constant $p_0$ and the local density is
larger than a randomly generated real number, $U_{01}$, uniform between 0 and 1, i.e.,
$p_0 / p(\vx|T) > U_{01}$. We set the value of $p_0$ to yield a subsample that is
half the size of the original data set.
Figure~\ref{fig:strat} shows the results for the previously selected sources based
on the smaller stratified subset.
The basic shape of the curves is practically the same in most cases,
only somewhat noisier but without systematics.
One exception is the blue galaxy at around $z=0.1$, where the
subsampling somewhat amplifies the effect of the large wall in SDSS at $z=0.08$.
The blue galaxy at the highest redshift is essentially unchanged
(except for the part at the lowest redshift where the density in magnitude space is larger to start with)
because the stratified sampling (by construction) has no effect on its already very sparse
neighborhood.

Optimal sampling is difficult to achieve. In fact, it is difficult even to define.
In addition to the photometric uncertainties,
the desired resolution of the physical quantities
also sets limits on the sampling frequency.
This is prominent in the case of degenerate regions where an extended part of the physical parameter
space is cramped into a small volume of observables.
Simulations built on realistic SED models can help cross-check
these factors, and evaluate the performance of the estimator
ahead of time. In the ideal case, one would
create stratified training sets in the space of the physical parameters instead of
the observables, which should be more feasible in the near future with improved
spectral modeling \citep[e.g.,][]{cb07}.

\subsection{Empirical Priors}

The distribution of sources in a training set may be artificial
and, as we just argued, should be optimized for coverage with a
practical upper bound on the density tuned to the photometric inaccuracies
and source diversity.
However, the distribution in the query set is often physical
and can be used to derive an empirical prior for our model.
The basic observation
is that the density of sources in the query set, $p(\vy|Q)$,
should match the predicted density of the model, $p(\vy|M)$.
The latter is calculated for any prior as the convolution,
\begin{equation}
p(\vy|M) = \int\!\!d\vt\,p(\vy|\vt,M)\,p(\vt|M)
\end{equation}
If we substitute $p(\vy|Q)$ measured from the sources
on the left-hand side of the equation, the only unknown is
the prior, which we can solve for using
the elegant deconvolution technique of \citet{richardson72} and \citet{lucy}.

To see why a physically sensible prior is important, let us
consider the density of sources in the training set within
the window function. Since the density is proportional to the
product of the underlying $p(\vx)$ density and the selection function,
\begin{equation}
p(\vx|T) \propto p(\vx)\,P(T|\vx)
\end{equation}
a significant volume of the window function
is not sampled by the training set,
where $p(\vx)$ is zero. Without a reasonable prior, the mapping
$p(\vx|\vy_q,M)$ could yield wrong weights for unphysical observables
in the summation of equation~(\ref{eq:pz}).
Hence, any model needs some physical information.
Even if one is hesitant to take the empirical prior at face value,
the domain of the model parameters should be carefully considered.
In case of template fitting,
this happens implicitly, even if not optimally, via the selection
of the set or manifold of template spectra,
but can be also done for even the empirical algorithms.

\section{Conclusions}   \label{sec:sum}

Starting from first principles of Bayesian probability theory,
we built a description and obtained the solution of the generic
photometric inversion problem, where the physical properties of
sources are constrained based on observational measurements.
The new approach yields a formalism that encapsulates
the field of photometric redshift estimation, and contains
the traditional methods as special cases. In our systematic
analysis of the mathematical problem, we put previous techniques
in context and pointed out the directions for improvement in each.

The proposed extensions to the current methods
represent significant progress in more respects.
We avoid the common assumption of the physical properties being a single-valued
function of the observables by treating their relation in a more general
way. 
Thus the formalism is not prone to fail in regions, where the data sets are
degenerate.
We showed how to estimate the corresponding probability density of this
relation.
In addition, the uncertainties of the observables are propagated all the way
to the results via explicit modeling of the accuracies. We discussed various
aspects of the modeling from the simplest empirical case to the application
of SEDs.

This general framework allows for the construction of novel, more advanced methods that
combine the attractive qualities of empirical and template-fitting algorithms. One can
build empirical estimators based on training sets that have different
observables from the query set, e.g., $U\!J\!F\!N$ photometry to $ugriz$, via SED modeling.
We can improve the methods by creating more and more realistic models that
include, for example, the strengths of the emission lines in galaxies \citep[following][]{zsuzsa}
and their inclination angles \citep[based on][]{yip_incl}
among the model parameters
to properly marginalize over the nuance parameters for a more reliable mapping
of the observables.

The current limitations 
come from the lack of good understanding of the photometric uncertainties.
From previous studies, we know that the flux measurements in
various passbands are correlated, yet,
most catalogs only quote errors on the individual fluxes.
For more precise scientific measurements via tighter photometric constraints,
we need better photometric error models in the future.
Upcoming survey telescopes will observe all sources multiple times,
hence will be able to get a better handle on the errors
and their covariances.
Understadning these systematics is probably one of the highest priority tasks in the preparation
for the upcoming era of photometric science.

The proper solution of the generalized photometric inversion problem
may be straightforward on paper,
but efficient implementations of realistic models with appropriate priors
involve many
advanced concepts in statistics, and can only be built on the most
recent and on-going developments in computer science, e.g., multi-dimensional indexing in databases.
Even then the computations are not trivial to carry out, and have significantly higher
demand for compute power than previous methods. The immediate future work
is to have such a unified framework developed and ready for the next generation
imaging surveys.

\acknowledgements %
The author is grateful to Alexander S. Szalay, Istv\'an Csabai and Andrew J.
Connolly for originally introducing the problem, and for the countless
inspiring discussions on various aspects of the topic over the course of many
years ever since.
The presented work has grown out of their collaboration on Hubble Space Telescope
and Sloan Digital Sky Survey data.
Also the author thanks Dongryeol Lee and Alexander G. Gray of the FastLab Team at the Georgia
Institute of Technology for making their dual-tree KDE implementation available
for this work, and for their continuing support to the astronomy community.
This study was supported by the Gordon and Betty Moore Foundation via GBMF 554.

\end{document}